# Mitigation of substrate coupling effects in RF switch by localized substrate removal using laser processing

Arun Bhaskar[1,2], Justine Philippe[1], Etienne Okada[1], Flavie Braud[1], Jean-François Robillard[1], Cédric Durand[2], Frédéric Gianesello[2], Daniel Gloria[2], Christophe Gaquière[1], Emmanuel Dubois[1]

[1] Univ. Lille, CNRS, Centrale Lille, Univ. Polytechnique Hauts-de-France, Yncréa Hauts-de-France, UMR 8520 - IEMN, F-59000 Lille, France
[2] STMicroelectronics, Crolles 38926, France

Corresponding author: Emmanuel Dubois (e-mail: emmanuel.dubois@isen.iemn.univ-lille1.fr).

**ABSTRACT** With the evolution of radio frequency (RF)/microwave technology, there is a demand for circuits which are able to meet highly challenging RF frontend specifications. Silicon-on-insulator (SOI) technology is one of the leading platforms for upcoming wireless generation. The degradation of performance due to substrate coupling is a key problem to address for telecommunication circuits, especially for the high throw count switches in RF frontends. In this context, a novel technique for local substrate removal is developed to fabricate membranes of mm-sized RF switch which allows for total etching of silicon handler. RF characterization of membranes reveal a superior linearity performance with lowering of 2$^{nd}$ harmonic by 17.7 dB and improvement in insertion losses by 0.38 dB in comparison with High-Resistivity SOI substrates. This improvement leads to a significant increase in frontend efficiency. These results demonstrate a new route for optimization of circuit performance using post-fabrication substrate processing techniques.

## I. INTRODUCTION

RF frontends are continuously evolving to support newer generations of communication standards. With each subsequent generation, different system specifications pertaining to losses, linearity and noise become more stringent. Therefore, RF designers are continuously in need of solutions to be able to meet these challenging specifications. RF switches are important components of the RF frontend as they allow to selectively propagate the signal through transmit/receive pathways using antenna-swapping, power and diversity switches to accommodate the increasing number of communication standards. There is a demand for high throw count switches in order to be able to support integration of different communication standards like GSM, TD-SCDMA, WCDMA and LTE within the same chip [1]. Another example of the demanding use of switches is given by carrier aggregation (CA) which is a key feature of the LTE-advanced and 5G standards. CA imposes extremely tight new specifications in terms of operational bandwidth, insertion loss, isolation, linearity and power handling [2]. Traditionally, GaAs p-HEMT and Silicon-on-Sapphire technologies have been preferred for switch design due to their superior substrate isolation [3][4] which was not possible on silicon. The introduction of high-resistivity SOI with substrate resistivity greater than 1 kΩ.cm was a breakthrough for flourishing of RF technology on SOI. A large number of switch designs have been reported ever since and a few examples are reported in [5]–[8]. The innovation in SOI has led to the development of two types of wafers: High-Resistivity SOI (HR-SOI) and Trap-Rich SOI (TR-SOI). In the former, despite a bulk substrate resistivity greater than 1 kΩ.cm, a parasitic surface conduction layer is formed at the interface between the buried oxide (BOX) and the silicon (Si) handler. This thin layer of locally enhanced resistivity contributes to substrate related losses and non-linearity. In the latter, this effect is mitigated by the introduction of a polysilicon layer naturally rich in trap states which strongly pin the potential and prevents the formation of a parasitic conduction layer. The trap rich layer greatly improves the loss and linearity figures of circuits as it has been shown in [9]–[11]. Despite significant improvements in SOI wafer technology, substrate coupling still occurs and degradation in performance is not negligible especially for bulky circuits like RF power switch. Improving the performance of switch has profound implications for the whole RF frontend. Cutting the switch losses by 0.1 dB can improve the frontend Tx efficiency by 1% and this improvement in efficiency takes around two years [12]. Substrate coupling increases the overall capacitance of the OFF branches which leads to losses and a degraded $R_{on} \times C_{off}$ figure-of-merit (FoM) [13]. Additionally, the parasitic substrate impedance is a source of non-linear behavior of the switch [14]. Thus, it is desirable to mitigate the substrate effects as shown in previous works where removal of handler substrate has proved to offer multiple benefits for different kinds of RF circuits [15]–[20]. In the aforementioned studies, handler substrate is replaced with another host substrate, plastic or glass, with reduced losses when compared to silicon. While there are improvements, the fabrication processes are cumbersome and substrate effects are not brought to a minimum due to the presence of the host substrate attached to the BOX by means of a bonding material.

In this work, we cross a distinctive milestone in substrate optimization. We report for the first time to our knowledge a methodology to locally suspend mm-sized RF devices/circuits with the silicon handler completely removed. After processing, only a few micrometers thick membrane remains in place consisting of the BOX, the active silicon layer and the interconnection network, which represents the ultimate minimal layer stack beyond which the electrical functionality cannot be preserved. The local suspension methodology is based on the use of a focused laser beam in order to perform a precise ablation to control the extracted volume of silicon and the roughness of the remaining material. For this purpose, a pulsed laser source is used to take advantage of the fascinating light-matter interaction mechanisms associated to a focused beam [21]. For instance, in the femtosecond pulse regime and for a fluence close to the ablation threshold, non-linear absorption is responsible for laser micromachining of silicon even in near infrared for which silicon is transparent under ordinary conditions [22]. Femtosecond laser processing is therefore utilized to etch handler silicon locally which fully eliminates substrate parasitics leading to enhanced RF circuits. A SP9T switch design is studied with a focus on improvements obtained at sub-6 GHz frequencies which is of interest for 5G applications [23].

The paper is organized as follows. Laser assisted substrate removal process is briefly described in section II. To quantitatively study substrate coupling in terms of losses and linearity, RF characterization of isolation structures is reported in section III. Finally, the performance of RF switch membranes are compared with HR-SOI and TR-SOI substrate types in section IV.

## II. LASER ASSISTED SUBSTRATE REMOVAL

The key step for the fabrication of membranes of switches is femtosecond laser ablation. A schematic explanation of the process is given in Fig. 1. A focused laser beam is scanned in a serpentine fashion as shown in Fig. 1b to ablate material within the area covered by the laser scan. Laser ablation is not uniform and ablated surfaces have finite surface roughness. Hence, laser ablation alone is not sufficient to remove material up to the BOX as such an attempt results in oxide layer etching in certain local areas. Laser ablation is therefore performed until remaining thickness of silicon is few tens of microns. The ablated cavities are then subject to xenon difluoride ($XeF_2$) vapor etching step. $XeF_2$ is highly selective etchant of Si over $SiO_2$ with a selectivity ratio of 1000:1. The unablated regions of the die are protected by using an etch-protect layer (Ajinomoto GX-T31) and it is placed in an etching chamber. Etching is performed over multiple cycles. In each cycle, $XeF_2$ gas is introduced into the chamber at pressure of 2-3 Torr for a time period of $10 - 15$ s and $XeF_2$ is vented out at the end of each cycle. About 25-100 cycles of etch are needed depending on thickness of silicon to be etched to completely remove the handler silicon in the ablated area to obtain membranes of circuits suspended on the BOX. The laser ablation process, also referred to as FLAME (Femtosecond Laser Assisted Micromachining and Etching), is fully detailed in [24] [25]. A high removal rate of up to 8.5 x $10^6$ $\mu m^3$ $s^{-1}$ is obtained by optimized laser milling parameters. The process is designed to retain handler silicon under RF pads to support the mechanical forces encountered during probe contact to the pads. The thickness of handler silicon under the pads is also reduced to few tens of microns at the end of the process.

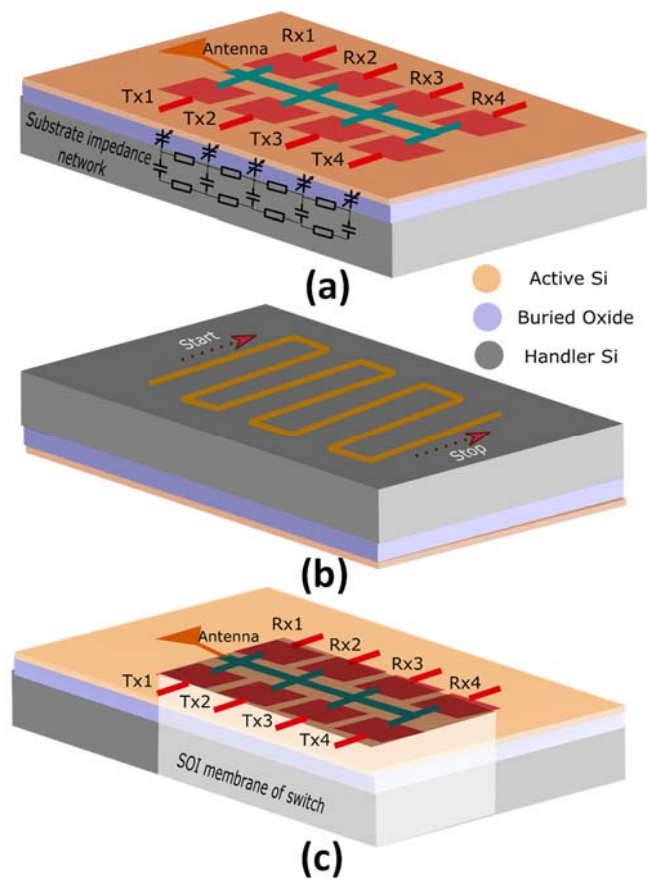

**FIGURE 1.** (a) Illustration showing the presence of substrate parasitics in RF switch (b) Trajectory of laser beam in the milling process to locally remove silicon from the back side (c) SOI membrane of switch showing complete local removal of handler silicon and consequently substrate parasitics.

## III. ISOLATION STRUCTURES

Isolation structures are specially designed to study and quantify in this work, the effect of substrate coupling in RF circuits. They also serve to demonstrate if membranes of circuit stacks can be reliably fabricated. A schematic illustration of these structures is shown in Fig. 2. They contain long interdigitated fingers separated by a defined spacing. The fingers start from metal 1 and they are contacted to the active silicon which is $p^+$ doped. The fingers are separated by shallow trench isolation (STI) as shown in Fig 2c. The interdigitated structures are connected to RF pads on either side which allows 2-port RF measurements. When RF signal is sent from one side, coupling takes place both through the layers above the BOX as well as through the substrate.

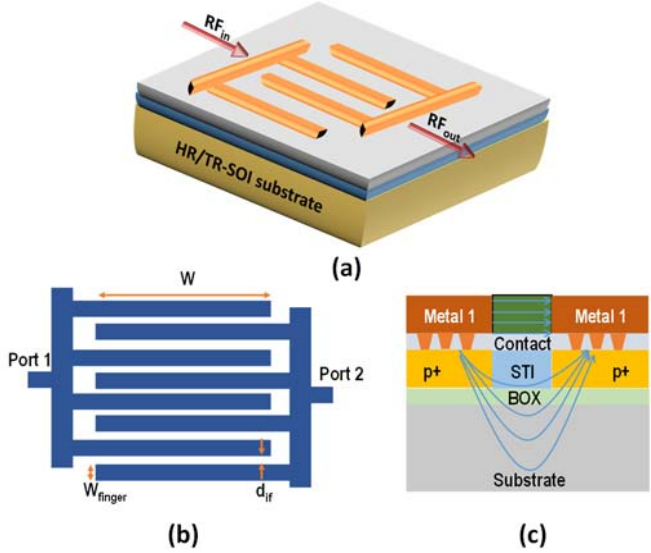

FIGURE 2. (a) 3D view of isolation structures showing interdigitated fingers (b) Top view (c) Cross-sectional view

TABLE 1. Dimensions of the studied isolation structures

| Reference | No. of Fingers | $d_{if}$ (μm) | W (μm) | $W_{finger}$ (μm) |
|---|---|---|---|---|
| ISO-A | 6 | 18.3 | 269.7 | 10.7 |
| ISO-B | 8 | 11.3 | 269.7 | 10.7 |

Two versions of the isolation structures are characterized: ISO-A and ISO-B. The different dimensions of the isolation structures are listed in Table 1. The difference between the two structures lies in the number of fingers and spacing between them. ISO-A has fewer fingers with higher spacing while ISO-B has higher number of fingers with smaller spacing. Two-port S-parameter characterization is performed over a frequency range of 20 MHz to 26 GHz. The measurements are performed using a Rohde & Schwarz ZVA 67 vector network analyzer and Infinity probes (GSG type) from Cascade Microtech are used to make contact with the RF pads on the die. The substrate is also biased through the chuck. S-parameter measurements are performed for different substrate bias conditions ranging from -2.5 V to +2.5 V with a step of 0.1 V. The isolation between the two ports characterized by $S_{21}$ (dB) is the parameter of interest. Lower value of $S_{21}$ (dB) corresponds to better isolation between the two ports.

Both HR-SOI and TR-SOI substrate types are characterized prior to substrate removal. $S_{21}$ values before laser processing for different frequencies plotted as a function of chuck bias are shown in Fig. 3. The HR-SOI substrate shows a strong dependence of $S_{21}$ on chuck bias clearly noticeable for MHz frequencies. This is because of the non-linear MOS like behavior of the substrate capacitance for which lower frequencies allow charge carriers to respond rapidly enough to the electric field. For frequencies < 2 GHz, $S_{21}$ is lower for positive chuck bias. At frequencies above 2 GHz, the transmission coefficient $S_{21}$ remains flat as the charge dynamics of majority carriers is profoundly affected by dielectric relaxation [26]. The dielectric relaxation time ($\tau_d$) is given by the product of the dielectric constant and the resistivity of the substrate. For the sake of illustration, $\tau_d$ is typically equal to ~2 ns, corresponding to a frequency of ~0.5 GHz, for a high resistivity substrate of 2 kΩ.cm. Under these conditions, the transport of majority carriers is impacted by an inertia effect which no longer allows carriers to respond fast enough to a high frequency excitation. This effect explains the flattening of $S_{21}$ at high frequencies in the HR-SOI substrate case.

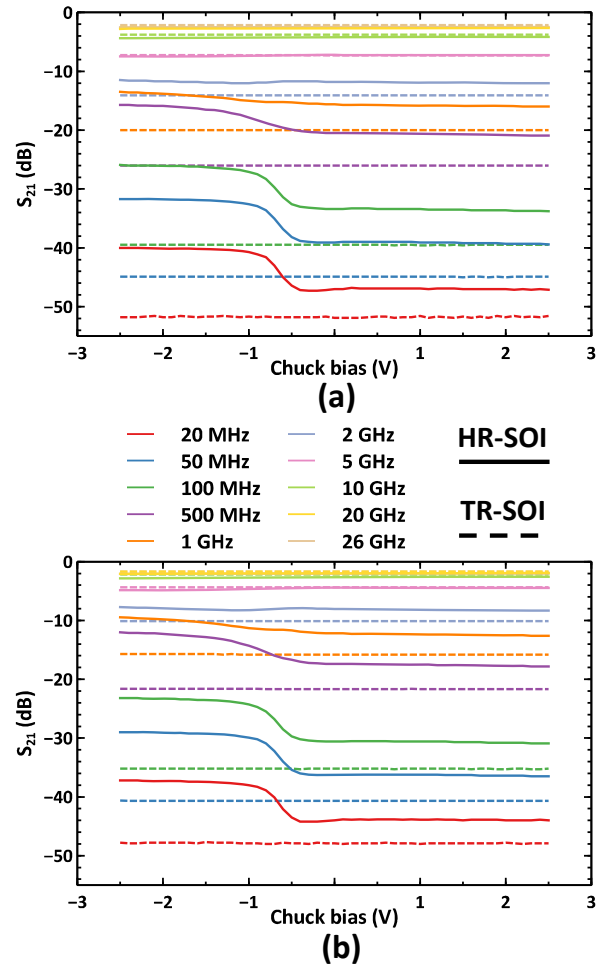

FIGURE 3. $S_{21}$ for two different isolation structures prior to substrate removal on both HR-SOI and TR-SOI substrate types (a) ISO-A (b) ISO-B

On the other hand, for TR-SOI substrate, the $S_{21}$ response at a given frequency remains flat regardless of the chuck bias. This behavior is assigned to the elimination of the interface conduction layer and also to the strong pinning of the potential due to the presence of a trap-rich layer that totally screens out the electrostatics below the BOX/handler interface [10]. At frequencies ≥5 GHz, the $S_{21}$ curves are flat for both substrates. It can also be noticed that $S_{21}$ at a given frequency is higher for ISO-B as compared to ISO-A. This is because of higher number of fingers available for coupling and also due to a closer spacing between them. Substrate removal is performed on ISO-A and ISO-B for both substrate types. $S_{21}$ values are plotted after substrate removal in Fig. 4. $S_{21}$ curves for the two substrate types match after FLAME process. At all frequencies, the isolation values for the two substrates are equal. These results indicate that the FLAME process for suspending RF circuits works well and the physical integrity of the circuit stack is maintained after substrate removal. In ISO-A, there is a very small but noticeable difference in isolation between the two substrates. These minor differences occur because of small variations between targeted and actual area of silicon removal for the two cases.

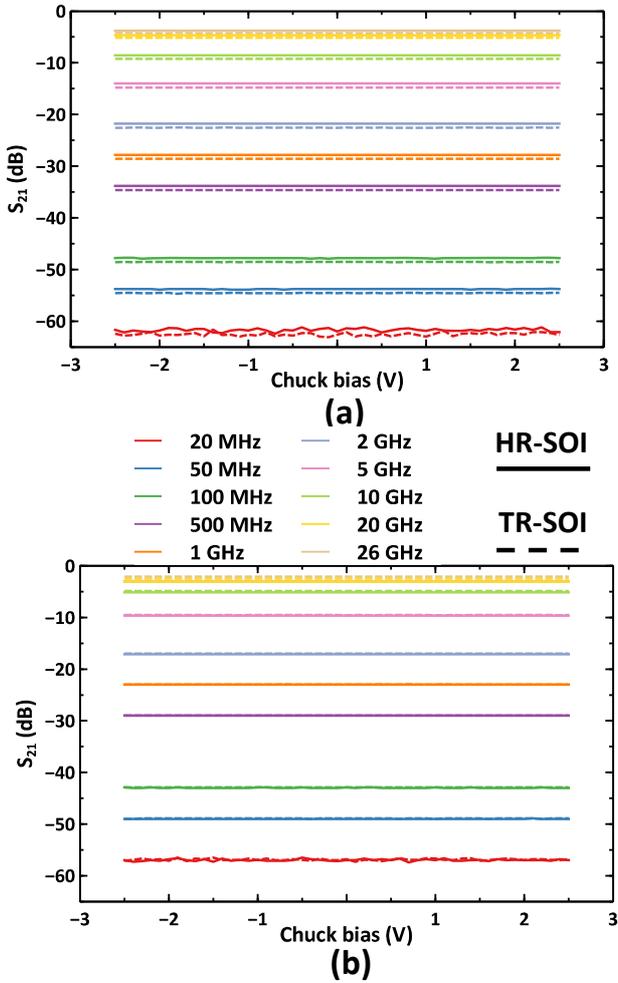

**FIGURE 4.** $S_{21}$ for two different isolation structures after substrate removal on both HR-SOI and TR-SOI substrate types (a) ISO-A (b) ISO-B

The etched area of the membrane is visualized using dual-light microscopy (DLM) where both the front side and back side are illuminated with varying intensities. The high intensity backlight through the cavity appears brighter in the microscope image outlining the area of membrane. The DLM images for ISO-A and ISO-B are shown in Fig. 5. It can be seen that bulb like features appear at some places around the outline of the membrane. These bulb-like features are a result of uneven local etching characteristic of $XeF_2$. Some places have an enhanced etch rate in the lateral direction as compared to the others. These features can be avoided by fine tuning the process in order to have fewer cycles of $XeF_2$ etching.

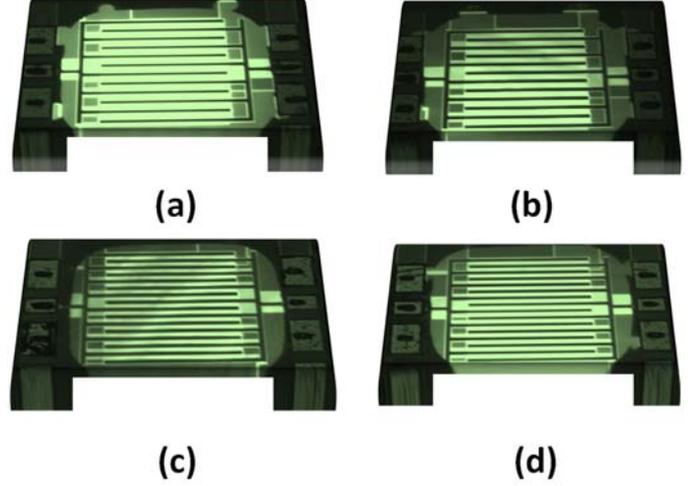

**FIGURE 5.** DLM images of isolation structures after substrate removal (a) ISO-A/HR-SOI (b) ISO-A/TR-SOI (c) ISO-B/HR-SOI (d) ISO-B/TR-SOI

In order to compare the isolation performance before and post FLAME process, the difference in isolation ($\Delta S_{21} = S_{21\ (HR/TR)} - S_{21\ (SR)}$) where SR refers to Substrate Removed is calculated for ISO-A and ISO-B and shown in Table 2. At MHz frequencies up to 100 MHz, the difference in isolation can exceed 20 dB for HR substrate for both isolation structures at negative chuck bias as seen in Fig. 3. The difference is smaller for TR-SOI substrate with values of 10.7 dB and 8.9 dB for ISO-A and ISO-B respectively at 20 MHz. At frequencies of operation of practical interest like 5 GHz, there is still a significant difference in isolation. The $\Delta S_{21}$ values (HR/TR) are 6.8/7.5 dB for ISO-A and 5.2/5.2 dB for ISO-B at 5 GHz and chuck bias of 0 V.

**TABLE 2.** $\Delta S21$ values depicting the difference in substrate coupling before and after removal of substrate at chuck bias of 0 V

| Reference | $\Delta S21$ (dB) (HR-SOI) | | $\Delta S21$ (dB) (TR-SOI) | |
|---|---|---|---|---|
| | 20 MHz | 5 GHz | 20 MHz | 5 GHz |
| ISO-A | 15 | 6.8 | 10.7 | 7.5 |
| ISO-B | 13 | 5.2 | 8.9 | 5.2 |

Thus, study of isolation structures demonstrates that coupling due to substrate parasitics can be considerably reduced by suspending the substrate for sub 6-GHz frequencies. In RF switches, substrate coupling occurs through the large number of source and drain fingers present within a transistor stack. A high density of fingers makes the substrate coupling significant. Another important observation is that any type of starting substrate can be used without affecting the final performance. Even low resistivity substrates can be used to achieve the same performance.

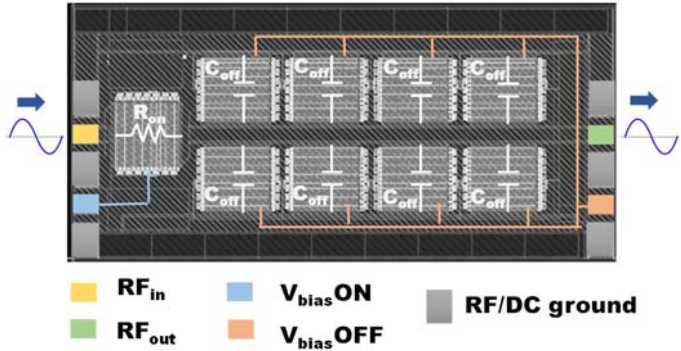

**FIGURE 6.** SP9T switch layout showing different pads in GSGSG configuration with ON/OFF state represented of each branch represented by $R_{on}/C_{off}$.

## IV. SP9T SWITCH DESIGN

An SP9T switch test structure is utilized in this work for demonstration of practical application. The switch is fabricated based on the 130nm H9SOIFEM process technology by STMicroelectronics [27]. The switch features 9 branches, each composed of a block comprising unitary MOS transistors assembled in parallel to lower the ON resistance and stacked in series to sustain high voltage. It constitutes a variation with respect the series-shunt architecture as described in [28]. This structure, which is conceptually more complex, has been designed to fully capture and better exacerbate the effects of loss and non-linearity related to the substrate. The following parameters pertaining to the ON state of the switch are measured: small-signal losses and linearity with respect to 2nd and 3rd harmonic (H2/H3) measurements. Accordingly, during characterization, of the 9 switch branches, one branch is in the ON state while rest of the branches are OFF as shown in Fig. 6. The OFF branches have a ground termination. To control the ON/OFF state of the different branches, the switch contains two DC bias pins. The body and gate are driven by these bias pins through a high series resistance for DC isolation [29]. The first pin is $V_{biasON}$ which puts one branch in the ON state by means of a positive bias voltage. A higher bias voltage gives a lower ON state resistance ($R_{on}$) for this branch. Two bias voltages for the ON branch are used in RF characterization: 2.5 V and 3.3 V. The second pin is $V_{biasOFF}$ which puts the remaining 8 branches in the OFF state by using a negative bias. For RF characterization, $V_{biasOFF}$ is set to -1.75 V. The negative bias controls the OFF state capacitance of the switch with more negative bias resulting in lower OFF state capacitance ($C_{off}$). Two implementations of the switch are characterized with different channel lengths: 180 nm and 220 nm. The layout area of each implementation is 1.26 x 0.86 mm$^2$.

## V. CHARACTERIZATION OF SP9T SWITCH AFTER HANDLER SUBSTRATE REMOVAL

### A. DC CHARACTERIZATION

DC characterization is performed to make a first analysis of performance of switch after substrate removal. Additionally, the impact of channel length on the switch ON-state characteristics is also highlighted. The gate bias for the ON branch ($V_{biasON}$) is varied and $I_d$-$V_d$ characteristics are recorded. The results are plotted in Fig. 7.

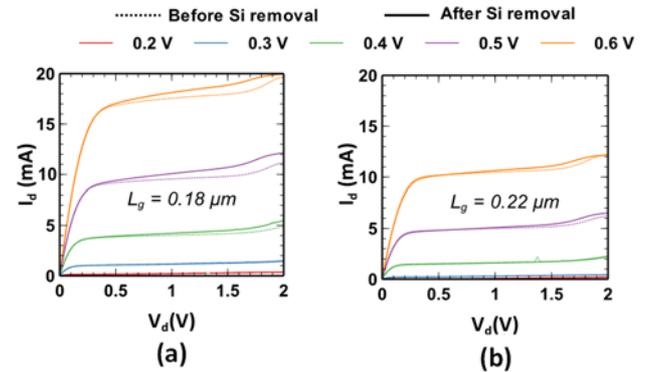

**FIGURE 7.** DC characteristics on TR-SOI substrate before and after silicon removal for two different channel lengths (a) 180 nm (b) 220 nm

No difference in drain current ($I_d$) can be observed after silicon removal in the linear regime and only a slight improvement in the saturation regime is seen that we assign to a variation of the threshold voltage. Good agreement in $I_d$-$V_d$ curves before and after substrate removal process suggests that the physical integrity of the circuit stack is maintained after substrate removal despite the large size of the membrane. Additional checks are performed to ensure that silicon is totally removed. The cross-sectional image of the switch cavity after laser processing and prior to XeF$_2$ etching is shown in Fig. 8a. After completion of XeF$_2$ etching, handler silicon is removed in portion outlined as grid in Fig. 8a. The substrate removal process is described in detail in [30]. To verify if silicon removal is complete, DLM images of the switch are taken as shown in Fig. 8b. Also images are taken with the die flipped to ensure that no traces of silicon are left in the area of the switch after XeF$_2$ etching. As expected, for the same applied gate voltage, the drain current is smaller for 220 nm channel length as compared to 180 nm channel length. Hence, a smaller ON state resistance and insertion loss is obtained for the 180 nm switch. Also, increasing the gate voltage reduces the resistance as observed by the slope of the IV curve in the linear region. The transistor operates in the linear region during the ON state. The use of a smaller gate length and a higher gate drive is therefore favorable to a lower ON state resistance. This

VOLUME XX, 2017    9

point is further corroborated when discussing S-parameter measurements in the following section.

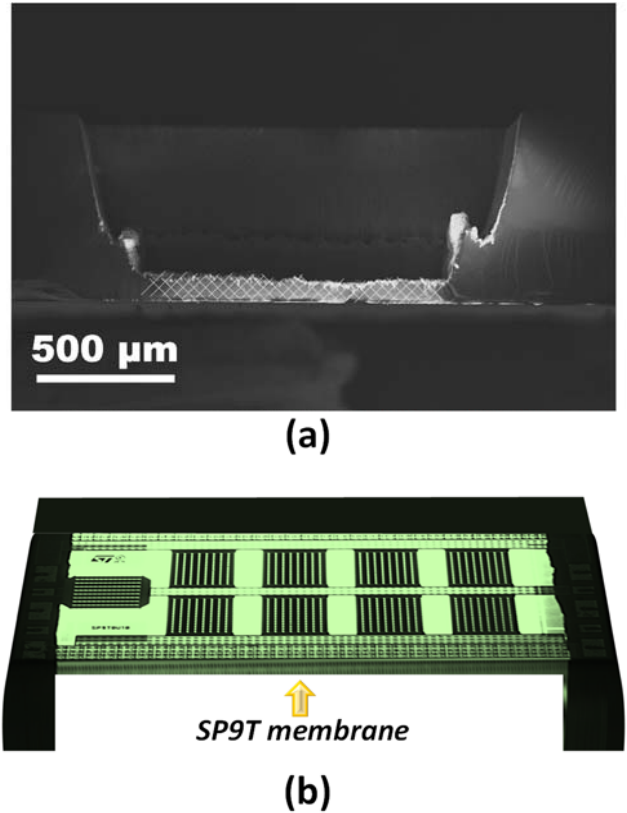

FIGURE 8. (a) SEM cross-sectional image of cavity fabricated using laser processing (b) 3D rendered backlight illuminated optical microscope image

## B. 2-PORT S-PARAMETER CHARACTERIZATION

S-parameter measurements are performed on the SP9T switch to determine the insertion loss and matching of the switch. The value of insertion loss can be taken approximately as -$S_{21}$ (dB) because the magnitude of the reflected signal on both ports are negligible compared to transmitted signal. The plots of the forward transmission coefficient $S_{21}$ are shown in Fig. 9.

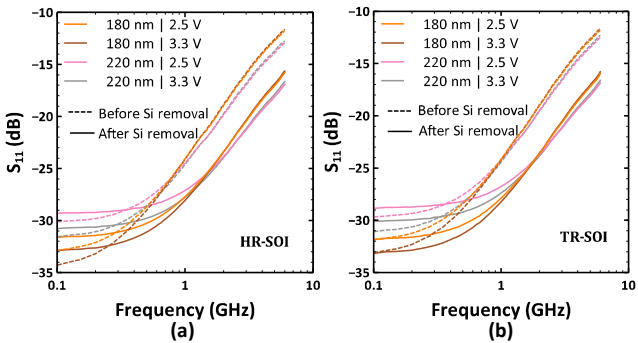

FIGURE 9. $S_{21}$ curves depicting the transmission coefficient of the switch before and after substrate removal for different channel lengths and $V_{biasON}$ values (a) HR-SOI (b) TR-SOI

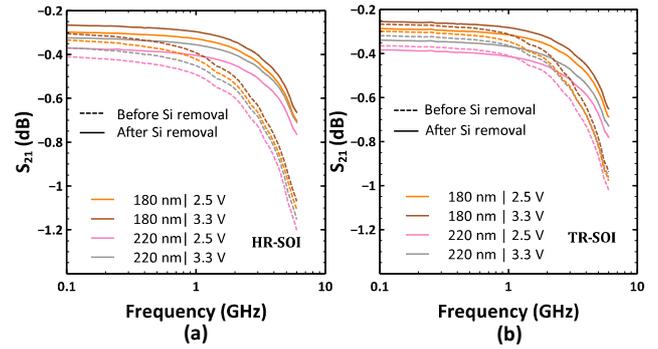

FIGURE 10. $S_{11}$ curves depicting the insertion losses of the switch before and after substrate removal for different channel lengths and $V_{biasON}$ values (a) HR-SOI (b) TR-SOI

Before substrate removal, a gate drive of 3.3 V is favourable for smaller losses as compared to 2.5 V. Also, insertion loss is lesser for channel length of 180 nm as compared to 220 nm. Both of these improvements can be attributed to reduction of on-state resistance of the transistor stack. The difference in on-state resistance can be clearly observed in the $S_{11}$ graphs shown in Fig. 10. At 100 MHz, $R_{on}$ is the most dominant component as parasitic capacitances are negligible at this frequency. A lower value of $R_{on}$ entails lesser reflection at the switch input port and hence lower value of $S_{11}$. It can be clearly seen that $S_{11}$ is lower for a smaller channel length and a higher gate voltage applied to the ON stack. As the frequency increases, the OFF stacks become more important in determining switch performance. It can be seen that at frequencies > 1 GHz, $S_{11}$ curves overlap for both bias conditions. At these frequencies, smaller value of $S_{11}$ corresponds to lower $C_{off}$. It can be seen that at 6 GHz, $S_{11}$ is lower for channel length of 220 nm as compared to 180 nm. However, losses are still marginally lower for channel length of 180 nm at this frequency because both $R_{on}$ and $C_{off}$ play a collective role in determining the losses. Another observation is that insertion losses increases at a faster rate as a function of frequency for 180 nm switch.

Substrate removal contributes to reduced insertion losses. The same trend is retained when the channel length and gate voltage are varied. The insertion losses remain about the same at MHz frequencies and reduce more notably at high frequencies. This is because at MHz frequencies switch insertion losses are dominated by $R_{on}$ which does not change appreciably after substrate removal. However, it can be noticed that for HR-SOI substrate a higher insertion loss even at MHz frequencies. This can be understood by looking at Fig. 3 where it was shown that substrate coupling is significantly larger at MHz frequencies for the HR-SOI substrate. It can also be observed that $S_{11}$ values at 100 MHz are higher after substrate removal. This can be attributed to the degraded contact which presents additional resistance in series with $R_{on}$. At 6 GHz, the average improvement (of the 4 cases) in $S_{21}$ is 0.38 dB for HR-SOI and 0.26 dB for TR-SOI. These improvements can be directly attributed to the significant reduction of $C_{off}$ in the OFF branches due to the elimination of substrate parasitics. This is validated by observing a ~4 dB reduction in $S_{11}$ for all 4 cases and both substrate types.

The reported improvements translate to significant enhancement in RF frontend efficiency. The frontend efficiency can be calculated by considering the Tx path consisting of a power amplifier (PA) in series with the switch [31].

$$\eta_{Tx} = \frac{P_{out}}{P_{in}} \quad (1)$$

$$P_{out} = \eta_{PA} P_{in} - P_{swloss} \quad (2)$$

Here $P_{in}$ is the input power to the PA, $P_{swloss}$ is the power dissipated in the switch, $P_{out}$ is the power delivered to the antenna, $\eta_{PA}$ is the power amplifier efficiency and $\eta_{Tx}$ is the frontend transmission efficiency. Consider a PA with 50% efficiency delivering a power of 1 W (30 dBm) to the switch input implying that $P_{in}$ is 2 W. The overall Tx efficiency in the case of 180 nm with $V_{biasON}$ of 3.3 V can be calculated at frequency of 6 GHz as 39.5%, 40.4% and 43% for HR-SOI, TR-SOI and substrate removed, respectively. This corresponds to an improvement of efficiency by 3.5% and 2.6% when comparing removed substrate to HR-SOI and TR-SOI, respectively. While the actual improvement in efficiency depends on the frontend architecture, this calculation shows the importance of switch losses for the overall efficiency of the RF frontend.

### C. LARGE SIGNAL HARMONIC DISTORTION MEASUREMENT

Harmonic distortion measurement has been performed to characterize the effect of substrate removal on linearity. The measurement is performed at a fundamental input frequency of 1.22 GHz while the 2nd and 3rd harmonics (H2/H3) are measured at the output at frequencies of 2.44 and 3.66 GHz, respectively. A low noise-floor setup as described in [29] is used for the measurement. The input power is swept at the fundamental frequency and the 2nd and 3rd harmonic power is recorded as shown in Fig. 11. As this is a sensitive measurement, higher sample size is used. The number of processed samples for second and third harmonics characterization is 4 and 5 for HR-SOI and TR-SOI substrate, respectively. As shown in Fig. 12, the dispersion of measured data is small before substrate removal and hence same could be expected for characterization after laser processing. However, after substrate removal the dispersion proves to be higher. This can be attributed to cumulative effects such as processing defects, improper probe contact due to pad degradation after probing multiple times, non-linearities introduced by the degraded pad and offset between targeted area and actual area of silicon removal.

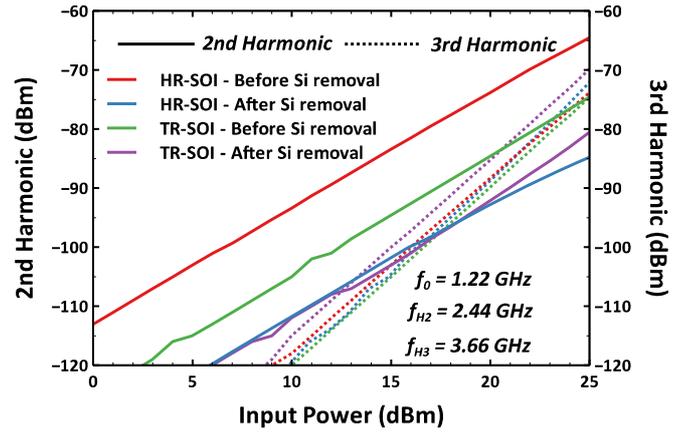

**FIGURE 11.** Harmonic distortion measurement shown for a sample with channel length 220 nm and $V_{biasON}$ = 2.5

Despite this measurement difficulty, Figure 12 shows very clearly a distinctive downward trend in harmonics with a level of reduction well beyond the dispersion range. In a switch, the second order non-linearities are dominated by $C_{off}$ and third order non-linearities by $R_{on}$ [14], [33]. With increase in gate voltage of the ON branch, both second and third harmonics levels reduce [34]. This trend is observed for both 180 nm and 220 nm gate length for both substrates before and after substrate removal. There is an overall improvement in H2 for both substrate types for both channel lengths and bias conditions after silicon removal. A detailed study of H2 for different substrate types has been done by Esfeh et al. [33]. For transistor stack in the OFF state, it has been shown that at a body bias of < -1.5 V, the substrate parasitics start to dominate over the device parasitics. It was also observed in case of isolation structures in section III that HR-SOI substrate has a non-linear behavior as compared TR-SOI. Therefore, prior to substrate removal, 2nd harmonics level is significantly higher for HR-SOI substrate. A removal of substrate leads to total elimination of the substrate parasitics and hence lesser harmonic distortion could be expected. Accordingly, there is an improvement in linearity for both HR-SOI and TR-SOI substrate types. The average improvement over two bias voltages for HR-SOI substrate is 17.7 dB and 16.1 dB for gate lengths 180 nm and 220 nm, respectively. For TR-SOI substrate, the average improvement is smaller compared to HR-SOI with 10.3 dB and 9.8 dB for gate lengths 180 nm and 220 nm, respectively. As for H3, similar performance is expected before and after substrate removal because $R_{on}$ does not change appreciably if handler substrate is removed. Hence, H3 values do not change appreciably for both substrate types. When looking at the average data, for HR-SOI substrate a slight improvement is observed while for TR-SOI substrate there is slight degradation. However, the large dispersion in data after substrate removal makes it difficult to ascertain both performance improvements and degradation in H3.

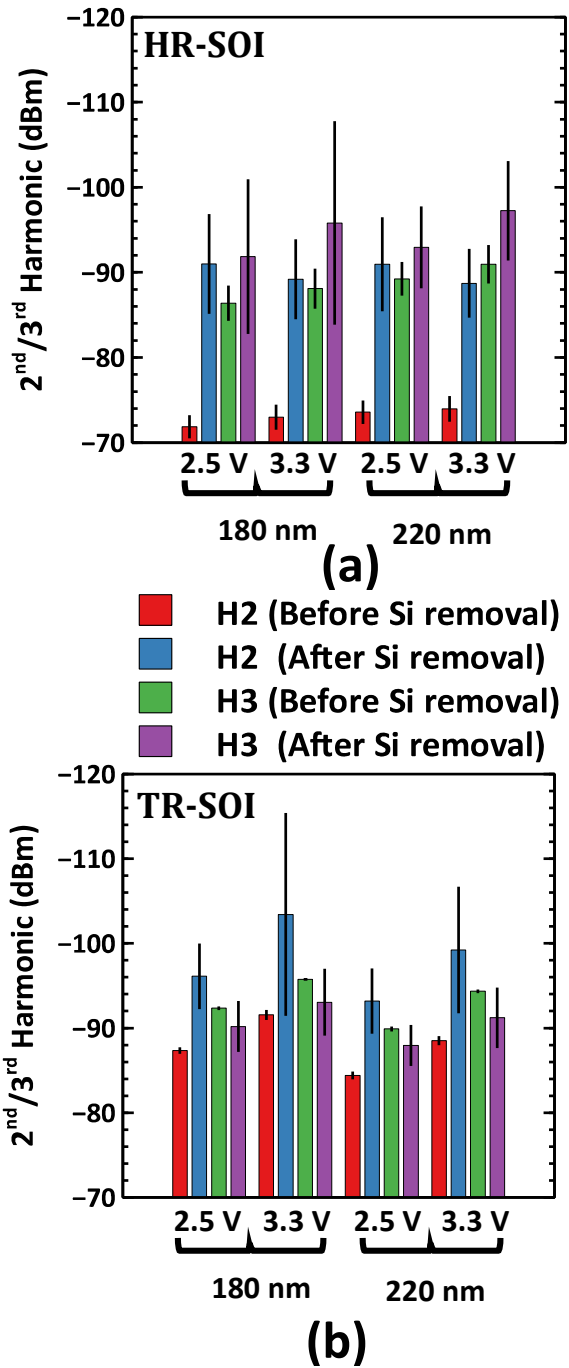

**FIGURE 12.** 2nd/3rd Harmonic levels for different channel lengths and bias conditions before and after substrate removal

## VI. CONCLUSION

A novel fabrication method using laser processing is presented which paves the way for complete handler substrate removal of SOI RF circuits. Membranes of isolation structures are first studied to delineate the importance of substrate in the performance of a RF circuit. Based on handler removal, it was shown that coupling is significantly reduced and any starting substrate can be used to arrive at the same final performance. A SP9T test structure was described and characterized to demonstrate the practical application of laser processing in real life circuits. The impact of eliminating substrate parasitics in the reduction of $C_{off}$ of the OFF transistor stacks was highlighted. Substantial gains in improvements are obtained. A reduction of insertion loss by 0.38 dB and 0.26 dB is obtained for HR-SOI and TR-SOI substrates, respectively. Calculations of frontend efficiency revealed that ~3% improvement can be reached by substrate removal. In addition to amelioration of losses, large signal linearity is also greatly enhanced. The second harmonic is suppressed further by up to 17.7 dB and 10.3 dB for HR-SOI and TR-SOI substrates, respectively. These results demonstrate the significance of local substrate removal methods to greatly improve the RF performance of the circuit and open up new avenues for RF designers.

### ACKNOWLEDGEMENTS
This work was supported by: i) the STMicroelectronics-IEMN common laboratory ii) the French government through the National Research Agency (ANR) under program PIA EQUIPEX LEAF ANR-11-EQPX-0025 and iii) the French RENATECH network on micro and nanotechnologies. The author would like to thank Jerome Lajoinie and Philippe Cathelin from STMicroelectronics for structure design and RF switch concept discussions.